# Thinning Energy Effect on the Fluctuations of Charged Particles Lateral Distribution Produced in Extensive Air Showers

Hassanen Abdulhussaen Jassim, A. A. Al-Rubaiee[*], Iman Tarik Al-Alawy

*College of Science, Department of Physics, Mustansiriyah University, Baghdad, Iraq*

*Email: dr.rubaiee@uomustansiriyah.edu.iq

## Abstract

In this work, effects of the extensive air showers (EAS) were described by estimating the lateral distribution function (LDF) at very high energies of various cosmic ray particles. LDF was simulated for charged particles such as the electron and positron pair production, gamma, muons and all charged particles at very high energies $10^{16}$, $10^{18}$ and $10^{19}$ eV. The simulation was performed using air shower simulator system (AIRES) version 2.6.0. The effect of primary particles, energies, thinning energy and zenith angle (θ) on the charged LDF particles-produced in the EAS was taken into account. The comparison of the estimated LDF of the charged particles such as the electron and positron pair production and muons with the simulated results by Sciutto and experimental results by Yakutsk EAS observatory gives a good acceptance at $10^{19}$ eV for 0̊ and 10̊ zenith angles.

## 1. INTRODUCTION

Extensive air showers are a cascade of electromagnetic radiation and ionized particles produced in the atmosphere through the interaction of primary cosmic ray with the atom's nucleus in the air producing a huge amount of secondary particles such as X-ray, electrons, neutrons, muons, alpha particles, etc [1]. In 1930, Pierre Victor Auger, the French physicist, discovered the EAS by producing more and more particles in the atmosphere [2]. The LDF of the charged particle in the EAS is a quantity required for observations of Earth's cosmic radiation, which are often derived from EAS observables [3]. The parameter used to describe the shape of the lateral structure density is the lateral shape parameter in the NKG function "Nishimura-Kamata-Greisen function"[4, 5]. EAS develops in a convoluted way as a combination of electromagnetic and hadronic showers. It is important to achieve a detailed numerical simulation of the EAS to infer the properties of the primary cosmic radiation, produced by it. The number of charged particles in ultra-high energy EAS may be enormous and may exceed $10^{10}$, so these processes require highly complex computing resources to understand and simulate them [6]. Since the shower growth is a complicated random process, Monte-Carlo simulation is often used to design atmospheric showers [7]. Among the many ways to simplify the problem and to reduce computation time, the thinning approximation is the most common entirety of the importance. It's essential idea is to track only a representative set of particles. While they are highly effective in calculations and provide the right values for observation on average, this method offers artificial fluctuations because the number of tracked particles decreases by several orders of magnitude. These artificial fluctuations are combined with natural fluctuations and thus reduce the precision of determination of physical parameters [8]. In 2007, Kuzmin studied no-thinning simulations of EAS and small-scale fluctuations at ground level [9]. In 2009, Bruijn studied statistical thinning with a full simulated of air showers at very high energies [10]. In 2015, Alex Estupiñan studied the achievement of the de-thinning method order to simulate EAS for high-energy cosmic rays [11]. Ivanov recently (in 2018), studied the distribution of the zenith angle of cosmic ray showers measured with the Yakutsk array and its application to the analysis of access trends in the equatorial coordinates [12].

The results of the current calculations have shown the effect of thinning energy on the fluctuations in the density of charged particles reaching the Earth's surface, such as the pair production of electron and positron, gamma, muons and all charged particles, by simulating the LDF carried out using the Monte Carlo AIRES system at ultrahigh energies $10^{16}$, $10^{18}$ and $10^{19}$eV. The estimated LDF comparison of charged particles such as the electron and positron pair production and muons with simulated results by Sciutto and Yakutsk EAS observatories gives good approval at $10^{19}$eV with thinning energies ($\epsilon_{th}=10^{-6}$ and $10^{-7}$) [13, 14].

## 2. LATERAL DISTRIBUTION FUNCTION

LDF of charged particles in the EAS is a significant amount of ground monitoring of cosmic radiations, through which most the cascade observables are deduced [15]. A study of EAS can be done experimentally on the Earth's surface, underground and in many mountains that rise by identifying some LDF quantities. i.e. the density of charged particles that originate in the EAS as a function of the basic distance of the shower core or in other words, the LDF is the shower structure of the cascade at different depths in the atmosphere [2]. The expression that is widely used to describe the LDF form is the NKG function that is presented through the forum[4]:

$$\rho(r) = \frac{N_e}{2*\pi*R_M^2} * C(s) * (\frac{r}{R_M})^{(s-2)} * (\frac{r}{R_M}+1)^{(s-4.5)} \qquad (1)$$

Where $\rho(r)$ is the particle density on the distance $r$ from the shower core, $N_e$ is the total number of shower electrons, $R_M$ =118 m is Molier radii, $s$ is the shower age parameter, and $C(s)$ is the normalizing factor of $0.366\, s^2 * (2.07-s)^{1.25}$ [16].

## 3. THINNING METHOD

The implementation of the thinning algorithm is used by simulating the shower on secondary particles if this condition is satisfied[11]:

$$E_\circ \epsilon_{th} > \sum_{j=1}^{n} E_j \qquad (2)$$

where $E_j$ is the secondary particle energy, $E_\circ$ is the energy of the primary particle and $\epsilon_{th} = E_j/E_\circ$ is defined as the level of thinning.

In this case, there is only one secondary particle $i$ can survive. The survival probability is:

$$P_i = E_i / \sum_{j=1}^{n} E_j \tag{3}$$

Otherwise, if the total amount of secondary particles $n$ is greater than the thinning energy threshold, i.e.:

$$E_\circ \epsilon_{th} < \sum_{j=1}^{n} E_j \tag{4}$$

then the secondary particle with energy below the thinning threshold will survive with a probability:

$$P_i = E_i / E_\circ \epsilon_{th} \tag{5}$$

## 4. RESULTS AND DISCUSSION

### 4.1 Simulating of LDF using AIRES system

AIRES is an acronym for AIR-shower Extended Simulations, which is defined as a set of programs and subroutines that are used to simulate EAS particle cascades, which initiated after interaction of primary cosmic radiations with high atmospheric energies and the management of all output associated data. AIRES provides a complete space-time particle propagation in a real medium, where the features of the atmosphere, the geomagnetic field, and Earth's curvature are adequately taken into account [13].

The thinning algorithm (statistical sampling step) is used when the number of particles in the showers is very large. The thinning algorithms used in AIRES are localized, i.e. statistical samples never change the average values of the output observables. There are many particles that are taken into account through simulations using the AIRES system such as: "electrons, positrons, gammas, muons, and all charged particles". The primary particle of the incident in the EAS may be a primary proton or iron nuclei or other primaries mentioned in the AIRES guidance document with a very high primary energy that may exceed $10^{21}$ eV [13].

Figure 1 shows the density of several secondary particles as a function of the distance from the shower axis that reaches the Earth's surface by AIRES simulation. The effect of primary particles (proton and iron), energies ($10^{16}$, $10^{18}$ and $10^{19}$ eV), zenith angles ($\theta = 0$, 10, 30 and 45 degrees) and the average of thinning energies ($\epsilon_{th}=10^{-3}$, $10^{-4}$, $10^{-6}$ and $10^{-7}$) on the density of charged particles produced in the EAS was taken into consideration. As shown in figure 1, the density of several secondary particles decreases with increasing distance from the shower axis. Finally, the statistical fluctuations of LDF of several secondary particles decrease while reducing the thinning energy.

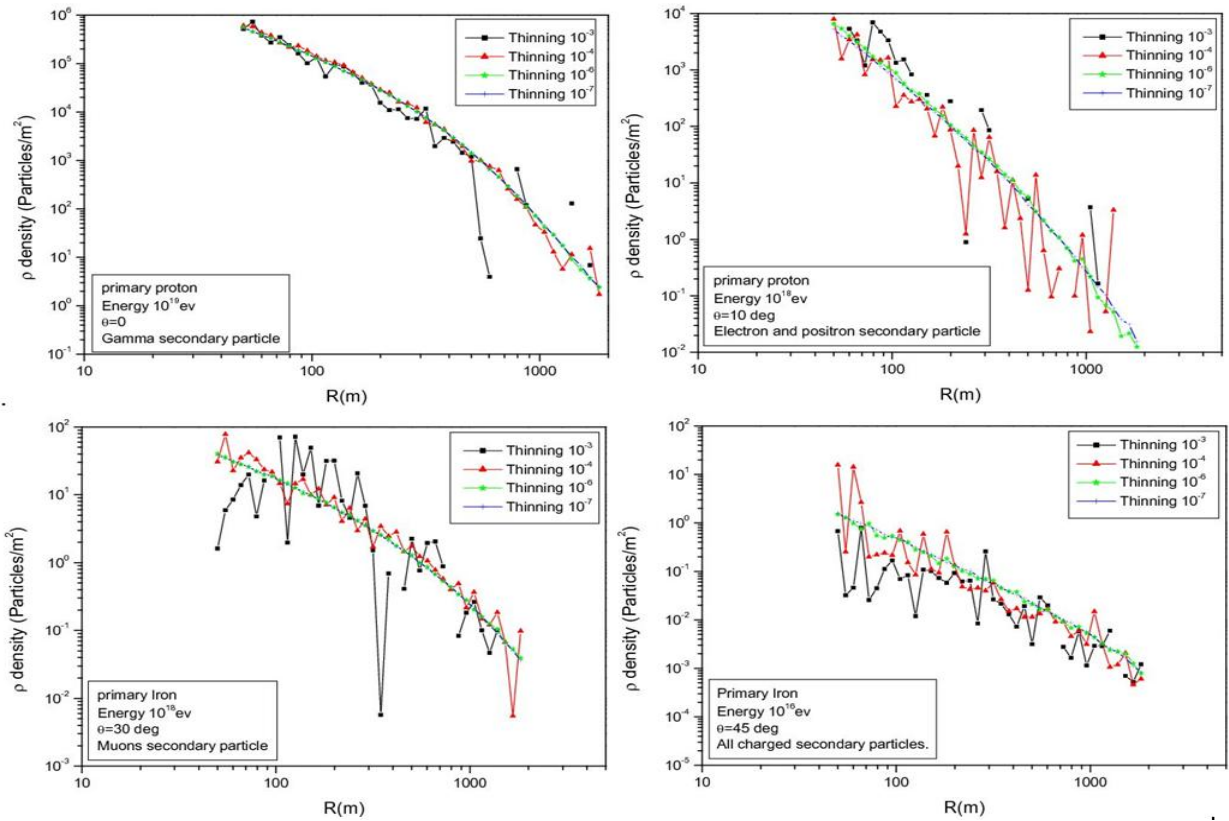

**Figure 1** The effect of thinning energy on secondary particle densities of primary p and Fe at different zenith angles (θ = 0°, 10°, 30° and 45°) and different energies ($10^{16}$, $10^{18}$ and $10^{19}$ eV).

*4.2 Comparison with the experience of Sciutto and Yakutsk Observatory*

Figure 2 demonstrates the comparison between the present results of LDF that performed by AIRES simulation (solid lines) with the results simulated by Sciutto (triangle symbols) [13]. This figure displayed good agreement between the secondary particles (electron and positron) and the muons particles initiated by the primary proton at energy $10^{19}$ eV with thinning energies ($\epsilon_{th}=10^{-6}$, $10^{-7}$) and the vertical EAS showers.

The Yakutsk EAS array studies the very high energy cosmic radiations, which occurs in the field of astrophysics, that is, an important area in physics. There are two main goals for construction of the Yakutsk EAS Observatory; the first is the elementary particles that verify the cascades that initiated by the primary particles in the atmosphere. The second goal is to reconstruct the astrophysical characteristics of primary particles such as: "mass composition, energy spectrum, intensity and their origin"[14].

Figure 3 shows the comparison between the present results and the experimental data obtained by the Yakutsk Observatory [14]. The curves in this figure displayed a good agreement for (electron and positron) and muons particles, which were initiated by primary proton at energy $10^{19}$ eV and a slanted EAS showers with θ =10°.

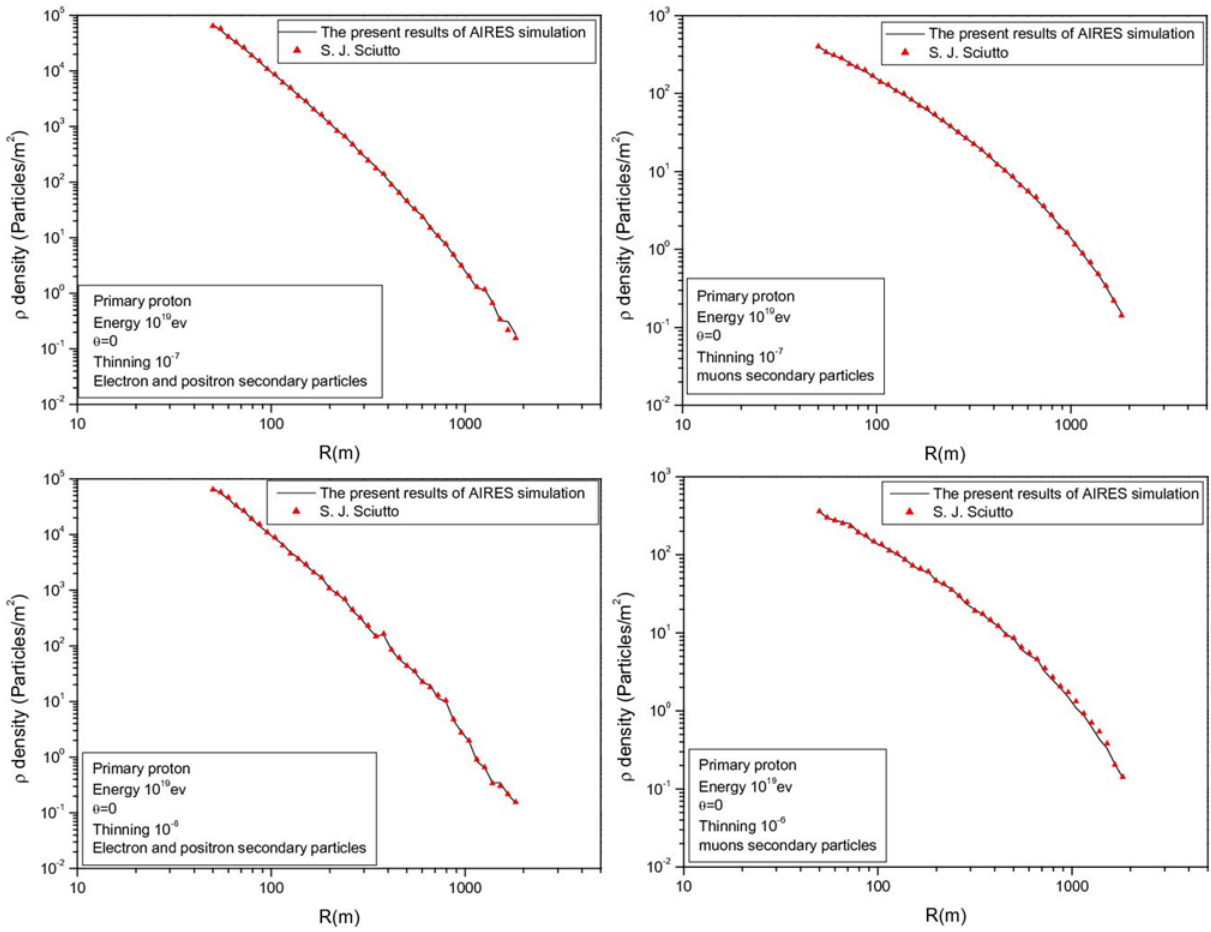

**Figure 2** Comparison between the present results of LDF simulation by the AIRES system with the results simulated by Sciutto for primary proton at $10^{19}$ eV with ($\epsilon_{th}=10^{-6}$ and $10^{-7}$) for secondary particles (electron and positron) and muons.

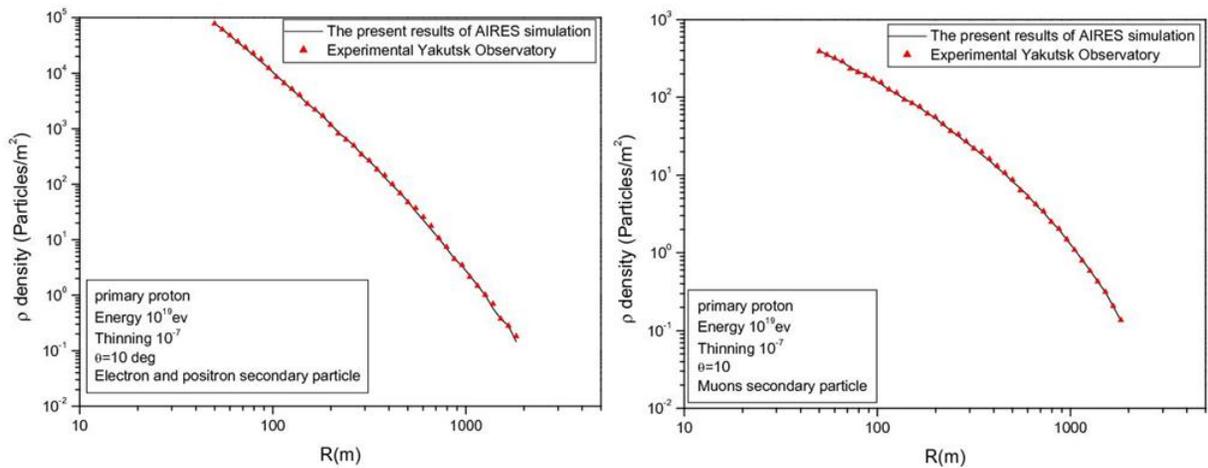

**Figure 3** Comparison between the present results of LDF simulation by the AIRES system with the experimental data obtained by Yakutsk Observatory for primary proton at $10^{19}$ eV for secondary particles (electron and positron) and muons.

## 5. CONCLUSIONS

In the present work, the lateral distribution function of charged particles using the AIRES system for two primary particles (proton and iron nuclei) was simulated in different ultrahigh energies $10^{16}$, $10^{18}$ and $10^{19}$ eV. The Simulation of lateral structure of the charged particle demonstrates the ability for distinguishing the primary cosmic ray particle and its energy. The statistical fluctuations of LDF of several secondary particles decrease with decreasing the thinning energy. An important feature of the present work is the creation of a library of lateral structure samples that can be used to analyze real EAS events that have been detected and registered in EAS arrays.

The introduced results using AIRES system are identified with Yakutsk experimental data, proving that AIRES provides an environment suitable for studying high-energy cosmic rays. Therefore, charged particles reaching the Earth's surface have many effects on weather, human health and other effects.